\documentclass[preprint,showpacs,preprintnumbers,amsmath,amssymb]{revtex4}
\usepackage{graphicx}
\usepackage{bm}

\begin{document}

\title{Particle transfer and fusion cross-section for
Super-heavy nuclei in dinuclear system}

\author{Wenfei Li$^{1}$, Nan Wang$^{3}$, Fei Jia$^{1,4}$, Hushan
Xu$^{1}$, Wei Zuo$^{1,5}$, Qingfeng Li$^{2,5,6}$, Enguang
Zhao$^{2,5}$, Junqing Li$^{1,2,5}$-\footnote{Corresponding author.
E-mail address: jqli@impcas.ac.cn}, and W. Scheid$^{5}$}
\vspace{5mm}
\affiliation{ \small  $^{1}$Institute of Modern
Physics, Chinese Academy of
Sciences, Lanzhou 730000, P.R.China,\\
\small $^{2}$Institute of Theoretical Physics, Chinese
Academy of Sciences, Beijing 100080, P.R.China,\\
\small $^{3}$Department of physics, college of science, Shenzhen University, Shenzhen 518060, P.R.China,\\
\small $^{4}$Graduate School of the Chinese Academy of
Sciences, Beijing 100039, P.R.China, \\
\small $^{5}$Institut fuer Theoretische Physik,
Justus-Liebig-Universitaet, 35392 Giessen, Germany,\\
\small $^{6}$ Frankfurt Institute for Advanced Studies (FIAS), Johann Wolfgang Goethe-Universit\"{a}t, D-60438 Frankfurt am Main, Germany
}

\date{June 1, 2005}

\begin{abstract}
Within the dinuclear system (DNS) conception, instead of solving
Fokker-Planck Equation (FPE) analytically, the Master equation is
solved numerically to calculate the fusion probability of
super-heavy nuclei, so that the harmonic oscillator approximation
to the potential energy of the DNS is avoided. The relative motion
concerning the energy, the angular momentum, and the fragment
deformation relaxations is explicitly  treated to couple with the
diffusion process, so that the nucleon transition probabilities,
which are derived microscopically, are time-dependent. Comparing
with the analytical solution of FPE, our results preserve more
dynamical effects. The calculated evaporation residue cross
sections for one-neutron emission channel of Pb-based reactions
are basically in agreement with the  known experimental data
within one order of magnitude.

{\bf keywords:} super heavy-nuclei; dinuclear system; driving
potential; master equation; complete fusion

\end{abstract}


\maketitle
%
\section{Introduction}
Since the nuclear shell model based on the Strutinsky
shell-correction method predicted that the next doubly magic shell
closure beyond $^{208}$Pb is at a proton number between Z=114 and
126 and a neutron number
N=184 \cite{Nil68,Mel67,Mos69,Fis72,Ran76}, a super-heavy nuclear
island of stability is expected, and the outstanding aim of
experimental investigation is the exploration of this region of
super-heavy elements. Up to now 16 new elements beyond fermium
(charge number Z=100) have been synthesized in the world, but
where is the center of the island is still an open question
\cite{Hof88,Oga99,Oga001,Hof20}. Furthermore, super-heavy elements
are extremely difficult to be synthesized because the formation
cross sections are very small, and the excitation functions are
very narrow. So a better understanding of the physics conception
on the super-heavy nucleus is very important. Several theoretical
models have been developed to describe the reaction dynamical
mechanism \cite{Smo99,She02,Fro84,Zag01,Che96,Ada97}. Among these
models, Adamian et al. have investigated the reaction mechanism of
the super-heavy element (SHE) formation in the concept of a
dinuclear system (DNS). In this model the formation is discussed
as a competition between quasi-fission and complete fusion, and
the cross sections are calculated including nuclear structure
effects. The model not only reproduces the experimental data quite
well, but also predicts the optimal projectile-target combination
as well as the optimal bombarding energy to form a certain SHE. It
is shown that the DNS model is a powerful tool to describe SHE
production, and is one of a few models so far which gives no
contradiction to available experimental data.

In the DNS model by Adamian et
al. \cite{Che96,Ada97,Ada98,Ant93,Ada972} it is considered that
after full dissipation of the collision kinetic energy, a DNS is
formed. The DNS evolves to a compound nucleus by nucleon transfer
from a light nucleus to a heavy one, and the
Fokker-Planck-Equation(FPE) is used to describe the diffusion
process. However, in this work \cite{Ada97,Ada98,Ant93,Ada972} a
Gaussian-type function solution of FPE is adopted with a harmonic
oscillator approximation of the potential energy surface. Or
Adamian et al. have used the Kramers-type expression solution
under the quasi-stationary approximation and the harmonic
oscillator approximation as well.  In fact the potential energy
surface of DNS including shell structure and even-odd effect
corrections deviates much from the harmonic oscillator form and
the approximation to it weakens the structure effect.  Recently
they have solved Master equation numerically for calculating the
charge, mass, and kinetic energy distributions of quasifission
products, describing the evolution of a dinuclear system in charge
and mass asymmetries and the decay of this system along the
internuclear distance.  But the coupling of the diffusion process
with relative motion is not considered \cite{Adac03,Dia01}.  The
dissipation of relative kinetic energy of heavy ions should be a
function of the interaction time of the DNS \cite{Wol78}; during
this time the nucleon transfer is coupled with the energy
dissipation, which is coupled with the angular momentum and the
dynamical deformation relaxations as well. To take these effects
into account within the DNS concept, instead of solving FPE
analytically, we have solved the Master equation numerically to
treat the nucleon transfer, so that the harmonic oscillator
approximation to the driving potential is avoided, and the nucleon
transfer process is coupled with the relative motion \cite{LWF03}.
Based on the fusion probability obtained from the numerical
solution of the Master equation, together with the calculations of
the survival probabilities, the evaporation residue cross sections
are obtained.

In section II the Master equation is introduced to calculate the
fusion probability of the DNS. The nucleon transition probability,
the rate of the quasi fission decay probability, the local
excitation energy and the driving potential of the system are
explained. In section III, for Pb-based one-neutron emission cold
fusion reactions and for compound nuclei from Z=106 to 118 (only
for even Z), the driving potentials of the DNS, the optimal
excitation energies, the fusion probabilities, the survival
probabilities, as well as the evaporation residue cross sections
are calculated. Our summary is given in section IV.
%

\section{Nucleon transfer in DNS concept }
\subsection{ The evaporation residue cross section in DNS
concept}
In the DNS concept the evaporation residue cross section
can be written as a sum over all partial waves J \cite{Ada98}
\begin{equation}\label{1}
  \sigma_{ER}(E_{cm})=\sum_{J=0}^{J=J_{f}}\sigma_c(E_{cm},J)
  P_{CN}(E_{cm},J)W_{sur}(E_{cm},J),
\end{equation}
where the partial capture cross section for the transition of the
colliding nuclei over the entrance barrier with the penetration
coefficient $T(E_{cm},J)$ at the incident energy of center of mass
$E_{cm}$ to form the DNS is given by
\begin{equation}\label{2}
  \sigma_{c}(E_{cm},J)=\pi{\lambda\hspace{-2.3mm}^-}^{2}(2J+1)T(E_{cm},J),
\end{equation}
where ${\lambda\hspace{-2.3mm}^-}$ is the reduced de Broglie
wavelength, ${\lambda\hspace{-2.3mm}^-}^{2}=\hbar^2 / (2\mu
E_{cm})$, with $\mu$ the reduced mass. $J_{f}$ is taken as the
value at which the contribution of the corresponding partial wave
to $\sigma_{ER}(E_{cm})$ becomes 0.01 times smaller than the
contribution of $J=0$ partial wave. For cold fusion reactions
leading to super-heavy nuclei, the values of $J_{f}\simeq 20 - 30
\hbar$. And $T(E_{cm},J)\simeq 0.5$ are chosen for energies
$E_{cm}$ near above the Coulomb barrier. The probability
$P_{CN}(E_{cm},J)$ of  the complete fusion is evaluated by
considering the fusion process as a diffusion of DNS in the mass
asymmetry $\eta=(A_{1}-A_{2})/A$, with $A_{1},A_{2}$ the mass
numbers of the DNS nuclei, and $A=A_{1}+A_{2}$. The nucleon
transfer is coupled with the dissipation of the relative kinetic
energy and the angular momentum, and the relaxation of colliding
nuclear deformations. The survival probability $W_{sur}(E_{cm},J)$
estimates the competition between fission and neutron evaporation
of the excited compound nucleus by the statistical model and
decreases much with increasing $J$, which determines the maximal
contributing $J_f$.

\subsection{ The Master equation}
In the DNS model \cite{Ada97,Ant93,Ada972} the dynamics has been
treated as a diffusion in mass asymmetry at the touching point to
the compound nucleus, and in the variable R of the relative
distance between the centers of the interacting nuclei, which may
lead to the quasi-fission. The analytical solution of the
Fokker-Planck equation \cite{Ada97,Ant93} or the numerical
solution of the Master equation \cite{LWF03,Dia01} are used to
describe the diffusion process. The nucleon transfer from the
light nucleus to the heavy one can be described by transport
theory which has been proved to be a successful tool for
investigating nucleon transfer in deeply inelastic collisions of
heavy ions
\cite{Nor74,Mor75,Hof76,Ran78,Wei80,Fro80,Fel87,Li88,Ant92,Ada94,Ant94}.

To solve FPE is convenient if it can be solved analytically. But
in this case the potential energy surface of DNS must be linear or
quadratic function of the relevant collective variables chosen to
treat the fusion process, and the harmonic oscillator
approximation of the potential energy surface is inevitable. To
avoid this approximation the Master equation is solved numerically
in order to treat the nucleon transfer in the present work.
Furthermore, in our investigation the nucleon transfer is coupled
with the relative motion and is considered as a time-dependent
process. The evolution of the DNS is not only a diffusion process
in the mass asymmetry at the touching point to the compound
nucleus, but also simultaneously in the variable R of the relative
distance between the centers of the interacting nuclei by decay
into the direction of increasing R, which may lead to the
quasi-fission of the DNS.  The fraction of the probability, which
goes to quasi fission, leaks out of the evolution system, so the
decay in R affects the motion of the system in $\eta$.  Let
$P(A_{1},E_{1},t)$ be the distribution function to find $A_{1}$
nucleons with excitation energy $E_{1}$ in fragment 1 at time t,
where $E_{1}$ is not considered as an independent variable but a
parameter supplied by the initial relative motion.
$P(A_{1},E_{1},t)$ obeys the following Master Equation(ME):

\begin{equation}\label{ma-quasi}
\frac{dP(A_{1},E_{1},t)}{dt}=\sum_{A_{1}^{'}}W_{A_{1},A_{1}^{'}}
[d_{A_{1}}P(A_{1}^{'},E_{1}^{'},t)-d_{A_{1}^{'}}P(A_{1},E_{1},t)]\\
-\Lambda ^{qf}_{A_{1},E_{1},t}(\Theta)P(A_{1},E_{1},t),
\end{equation}

where $W_{A_{1},A_{1}^{'}}=W_{A_{1}^{'},A_{1}}$is the mean
transition probability from a channel $(A_{1}, E_{1})$ to
$(A_{1}^{'}, E_{1}^{'})$, $d_{A_{1}}$ denotes the microscopic
dimension for the corresponding macroscopic variables. The
coefficient $\Lambda^{qf}_{A_{1},E_{1},t}(\Theta)$ is the rate of
decay probability in R, and will be described later. The sum is
taken over all possible mass numbers that fragment 1 may take(from
0 to $A=A_{1}+A_{2}$). The motion of the nucleons in the
interacting nuclei is considered to be described by the
single-particle Hamiltonian \cite{Nor74,Ayi76}

\begin{equation}\label{4}
  H(t)=H_{0}(t)+V(t)
\end{equation}
with
\begin{equation}\label{5}
H_{0}(t)=\sum_{k}\sum_{\nu_{k}}\varepsilon_{\nu_{k}}(t)
a_{\nu_{k}}^{\dagger}(t)a_{\nu_{k}}(t),
\end{equation}

\begin{equation}\label{6}
  V(t)=\sum_{k,k^{'}}\sum_{\alpha_{k},\beta_{k^{'}}}u_{\alpha_{k}\beta_{k^{'}}
}(t)a_{\alpha_{k}}^{\dagger}(t)a_{\beta_{k^{'}}}(t)=\sum_{k,k^{'}}V_{k,k^{'}}(t),
 \ k,k^{'}=1,2
\end{equation}

The quantities $\varepsilon_{\nu}(t)$ and $u_{\nu \mu}(t)$ denote
the single-particle energies and the interaction matrix elements,
respectively. The single-particle states are defined with respect
to the moving centers of nuclei and are assumed to be
orthogonalized in the overlap region. Therefore, the annihilation
and creation operators depend on time. The single-particle
interaction matrix element is parameterized by

\begin{equation}\label{7}
u_{\alpha_{k},\beta_{k^{'}}}(t)=U_{kk^{'}}(t) \{ exp
[-\frac{1}{2}(
\frac{\varepsilon_{\alpha_{k}}(t)-\varepsilon_{\beta_{k^{'}}}(t)}{\triangle_{kk^{'}}(t)}
)^{2} ] -\delta_{\alpha_{k},\beta_{k^{'}}} \},
\end{equation}
which contain five independent parameters.  These are the strength
parameters $U_{11}(t)$ and $U_{22}(t)$ for exciting a nucleon in
fragment 1 and 2, respectively, and $U_{12}(t)=U_{21}(t)$ for
transferring a nucleon between the fragments, and the
corresponding width parameters $\Delta_{11}(t)=\Delta_{22}(t)$ and
$\Delta_{12}(t)=\Delta_{21}(t)$. The strength parameters are taken
as($g_k=A_k/12$):

\begin{equation}\label{8}
U_{kk^{'}}=\frac{g_{1}^{\frac{1}{3}}g_{2}^{\frac{1}{3}}}
{g_{1}^{\frac{1}{3}}+g_{2}^{\frac{1}{3}}}\cdot\frac{1}
{g_{k}^{\frac{1}{3}}g_{k^{'}}^{\frac{1}{3}}}\cdot2\gamma_{kk^{'}}
\end{equation}

In our calculation $\Delta_{11}(t)=\Delta_{12}(t)=2MeV$, and the
dimensionless strength parameters
$\gamma_{11}=\gamma_{22}=\gamma_{12}=\gamma_{21}=3$ are
taken \cite{Nor74,Ayi76}. The transition probability reads:

\begin{equation}\label{9}
W_{A_{1},A_{1}^{'}}=\frac{\tau_{mem}(A_{1},E_{1},A_{1}^{'},E_{1}^{'})}
{\hbar^{2}d_{A_{1}}d_{A_{1}^{'}}}\sum_{ii^{'}}|\langle
A_{1}^{'},E_{1}^{'},i^{'} |V|A_{1},E_{1},i\rangle|^{2},
\end{equation}
where i denotes all remaining quantum numbers. The memory time is:

\begin{equation}\label{10}
\tau_{mem}(A_{1},E_{1};A_{1}^{'},E_{1}^{'})=(2\pi)^{1/2}\hbar\{
\langle V^{2}(t)\rangle\ _{A_{1},E_{1}}+\langle V^{2}(t)\rangle\
_{A_{1}^{'},E_{1}^{'}}\}^{-1/2},
\end{equation}
where $\langle $\ $ \rangle _{A_{1},E_{1}}$ stands for the average
expectation value with $A_{1},E_{1}$ fixed. Due to the dissipated
energy in the evolution process of the relative motion, the nuclei
are excited gradually. The excitation energy opens a valence space
of width $\Delta\varepsilon_{k}$ in the fragment k which lies
symmetrically around the Fermi energy surface. Only those
particles in the states within the valence space are active for
excitation and transfer. The averages in Eqs.(9) and (10) are
performed in the valence space:

\begin{equation}\label{11}
\Delta\varepsilon_{k}=\sqrt{\frac{4\varepsilon_{k}^{*}}{g_{k}}},
\varepsilon_{k}^{*}=\varepsilon^{*}\frac{A_{k}}{A},
g_{k}=\frac{A_{k}}{12},
\end{equation}
where $\varepsilon^{*}$ stands for the local excitation energy of
the system, and will be given below. There are
$N_{k}=g_{k}\Delta\varepsilon_{k}$ valence
states and $m_{k}=N_{k}/2$ valence nucleons in
$\Delta\varepsilon_{k}$. The dimension is
$d(m_{1},m_{2})=\left(\begin{array}{c}
N_{1}\\m_{1}\end{array}\right) \left(\begin{array}{c}
N_{2}\\m_{2}\end{array}\right) $.  The transitions for a proton or
neutron are not distinguished in the transition probability since
the ME is essentially restricted to one dimension. It is, however,
remedied by including the explicit proton and neutron numbers of
the isotopic composition of the nuclei forming the DNS in the
driving potential.  The averages in Eqs. (9) and (10) are carried
out by using the method of spectral
distributions \cite{Fre71,Cha71,Ayi74}. We obtain

\begin{equation}\label{12}
\langle V_{kk^{'}}V_{kk^{'}}^{\dag}\rangle=\frac{1}{4}
U_{kk^{'}}^{2}g_{k}g_{k^{'}}\Delta_{kk^{'}}\Delta \varepsilon_{k}
\Delta\varepsilon_{k^{'}}[\Delta_{kk^{'}}^{2}+\frac{1}{6}
(\Delta\varepsilon_{k}^{2}+\Delta\varepsilon_{k^{'}}^{2})]^{-1/2}
\end{equation}
and

\begin{equation}\label{13}
\tau_{mem}(A_{k},E_{k},t)=\hbar[2\pi/\sum_{kk^{'}}\langle
V_{kk^{'}}V_{kk^{'}}^{\dag}\rangle]^{1/2}
\end{equation}

According to Eq. (6) the transition probability of Eq. (9) can be
written as:

\begin{equation}\label{14}
\begin{array}{c}
W_{A_{1},A_{1}^{'}}(A_{1},E_{1};A_{1}^{'},E_{1}^{'})
=\frac{\tau_{mem}(A_{1},E_{1};A_{1}^{'},E_{1}^{'})}
{\hbar^{2}d_{A_{1}}d_{A_{1}^{'}}}\{[\omega_{11}(A_{1},E_{1},E_{1}^{'})+
\omega_{22}(A_{1},E_{1},E_{1}^{'})]\delta_{A_{1}^{'},A_{1}}\\
+\omega_{12}(A_{1},E_{1},E_{1}^{'})\delta_{A_{1}^{'},A_{1}-1} +
\omega_{21}(A_{1},E_{1},E_{1}^{'})\delta_{A_{1}^{'},A_{1}+1}\},
\end{array}
\end{equation}
where
\begin{equation}\label{15}
\omega_{kk^{'}}(A_{1},E_{1},E_{1}^{'})=\sum_{k,k^{'},A_{1}^{'}}|\langle
A_{1},E_{1},k|V_{kk^{'}}|A_{1}^{'},E_{1}^{'},k^{'}\rangle|^{2}
=d_{A_{1}}\langle V_{kk^{'}}V_{kk^{'}}^{\dag}\rangle.
\end{equation}

\subsection{ The local excitation energy and the driving potential of the system}
The local excitation energy is defined as the following:

\begin{equation}\label{16}
\varepsilon^{*}=E-U(A_{1},A_2)-\frac{(J-M)^{2}}{2\mathcal{J}_{rel}}
-\frac{M^{2}}{2\mathcal{J}_{int}},
\end{equation}
 where $E$ is the intrinsic excitation energy of the
composite system converted from the relative kinetic energy loss.
$M$ denotes the corresponding intrinsic spin due to the relative
angular momentum dissipation and $\mathcal{J}_{int}$ the
corresponding moment of inertia of the system. $J$ and
$\mathcal{J}_{rel}$ are the relative angular momentum and the
relative moment of inertia of the DNS, respectively. The
quantities $E$, $M$, $\mathcal{J}_{int}$, $\mathcal{J}_{rel}$
calculated for each initial relative angular momentum $J$, are
coupled each other due to the fragment deformation relaxation and
are functions of the evolution time t.

The driving potential energy for the nucleon transfer of the DNS
is:

\begin{equation}\label{17}
U(A_{1},A_2)=B(A_{1})+B(A_{2})-B(A)+
U_{C}(A_{1},A_2)+U_{N}(A_{1},A_2),
\end{equation}

where $B(A_{1})$, $B(A_{2})$, and $B(A)$ are the binding energies
of the fragments and compound nucleus, respectively, and are taken
from Ref. \cite{Moe95}, so that the shell and paring corrections
are included in them. The nuclear interaction energy can be
parameterized by the Morse potential as in Ref. \cite{Ada97}

\begin{equation}\label{18}
U_{N}(A_{1},A_2)=D(exp[-2\alpha\frac{R-R_{0}}{R_{0}}]-2
exp[-\alpha\frac{R-R_{0}}{R_{0}}]),
\end{equation}
where $D=2\pi a_{1}a_{2} R_{12}(10.96-0.8R_{12})$ (in MeV),
$R_{0}=R_{1}+R_{2}$, and
$\alpha=11.47+2.069R_{12}-17.32a_{1}a_{2}$ (dimensionless) are the
depth, minimum position, and inverse width of the potential,
respectively, $R_{12}=R_{1}R_{2}/R_{0}$ ($R_{1}$,$R_{2}$ are the
radii of the nuclei). $a_{1}, a_{2} \approx 0.54 - 0.59 $. $R$ is
the distance between the centers of nuclei. It is not taken as an
independent variable in our calculation, but as
$R=R_{1}+R_{2}+R_{d}$, where $R_{d}$ is chosen as the value which
gives the minimum value of $U_{C}(A_{1},A_2)+U_{N}(A_{1},A_2)$. If
the ground state deformations of the two touching nuclei are taken
into account, the Coulomb interaction of the deformed DNS,
$U_{C}(A_{1},A_2)$, must be calculated numerically.  For the
nuclear part of the potential, in Eq. (\ref{18}), nuclei are
assumed as spherical but shifted to a smaller relative distance
determined by the same distance between the nuclear surfaces as
the one which the deformed nuclei have.  In this manner the
deformation of the nuclei was simulated.  In principle, the
deformed nuclei can have different relative orientations. Some
averaging over the orientations of the nuclei has to be carried
out in the initial DNS, however, the orientation which gives rise
to the minimum interaction energy is in favor of the nucleon
transfer.  So the pole to pole orientation is chosen as the case
which gives rise to the minimum energy.

      The evolution of the DNS in the variable R of the relative
distance between the centers of the interacting nuclei will lead
to the quasi-fission of the DNS.  For a given mass asymmetry
$\eta$, the nucleus-nucleus interaction potential as a function of
R is:
\begin{equation}\label{vint}
V(A_{1},A_2,R)=U_{C}(A_{1},A_2,R)+ U_{N}(A_{1},A_2,R)+
U_{rot}(A_{1},A_2,R),
\end{equation}
where the Coulomb interaction $U_{C}$ is calculated numerically
and the nuclear interaction $U_{N}$ is calculated by Eq.(\ref{18})
as a function of R at each combination of the DNS. $U_{rot}$ is
the centrifugal potential. The nucleus-nucleus interaction
potential $V(A_{1},A_2,R)$ has a pocket as a function of the
relative distance R with a small depth which results from the
attractive nuclear and repulsive Coulomb interactions. The
probability $P(A_{1},E_{1},t)$ distributed in the pocket will have
the chance to decay out of the pocket with a decay rate
$\Lambda^{qf}_{A_{1},E_{1},t}(\Theta)$ in Eq.(\ref{ma-quasi}),
which can be treated with the one dimensional Kramers rate as in
Ref. \cite{Adac03}:
\begin{equation}\label{theta}
\Lambda ^{qf} _{A_{1},E_{1},t}(\Theta)=
\frac{\omega}{2\pi\omega^{B_{qf}}}
(\sqrt{(\frac{\Gamma}{2\hbar})^{2}+(\omega^{B_{qf}})^{2}}-\frac{\Gamma}{2\hbar})
exp(-\frac{B_{qf}(A_{1})}{\Theta(A_{1},E_{1},t)}),
\end{equation}
which exponentially depends on the quasi fission barrier
$B_{qf}(A_{1})$ for a given mass asymmetry $\eta$, and the
$B_{qf}(A_{1})$  measures the depth of this pocket. The
temperature $\Theta(A_{1},E_{1},t)$ is calculated by using the
Fermi-gas expression $\Theta=\sqrt{\frac{\varepsilon^{*}_{1}}{a}}$
with the excitation energy $\varepsilon^{*}_{1}$ given in
Eq.(\ref{11}), and $a=\frac{A}{12} MeV^{-1}$. $\omega^{B_{qf}}$ in
Eq.(\ref{theta}) is the frequency of the inverted harmonic
oscillator approximating the potential V in R around the top of
the quasi fission barrier. And $\omega$ is the frequency of the
harmonic oscillator approximating the potential in R at the bottom
of the pocket. They are determined by the local oscillator
approximation of the nucleus-nucleus potential energy. The
quantity $\Gamma$ denotes a double average width of the
contributing single-particle states, which determines the friction
coefficients:
  $\gamma_{ii^{'}}=\frac{\Gamma}{\hbar}\mu^{-1}_{ii^{'}}$, with
$\mu_{ii^{'}}$ the mass parameters. And $\Gamma\approx 2MeV$. From
our calculation, the extracted average values:
$\hbar\omega^{B_{qf}} \sim 2.0 MeV$, and $\hbar\omega \sim 4.0
MeV$.

\subsection{ The numerical procedure}
The distribution function $P(A_{1},E_{1},t)$ is calculated by
solving Eq.(\ref{ma-quasi}) numerically. During the nucleon
transfer process it is assumed that only one nucleon exchange is
preferential. Two or more than two nucleon exchange processes at
one time are negligible. These allow us to make a simplification
in Eq.(\ref{ma-quasi}) about the transition probability that
$W_{A_{1},A_{1}^{'}}$ are sharply and symmetrically peaked at
$A_{1}$ and only $W_{A_{1},(A_{1}-1)}$ and $W_{A_{1},(A_{1}+1)}$
are significant. The consequence is that only two terms, namely
 $A_{1}^{'}=A_{1} \pm 1$ remain in the summation of Eq.(\ref{ma-quasi}), so that the
difference equations corresponding to Eq.(\ref{ma-quasi}) become
tri-diagonal coupled algebraic equations.

The boundary of the distribution function $P(A_{1},E_{1},t)$ is
assumed as: $P(A_{1}<0,E_{1},t)=0$, and
$P(A_{1}>(A_{P}+A_{T}),E_{1},t)=0$, where $A_{P}$,$A_{T}$ are the
mass numbers of the projectile and the target, respectively. The
initial condition is $P(A_{1},E_{1},t=0)=\delta_{A_{1},A_{P}}$. In
all cases investigated, the time step interval $\Delta \tau$ is
taken to be from 0.05 to $0.1\times10^{-22}$sec. In the region
where the kinetic energy loss increases faster, the transition
probability changes also rapidly, the $\Delta\tau$ should be
smaller. Throughout the evolution process the normalization of the
distribution function must be preserved at the condition of the
decay rate $\Lambda^{qf}_{A_{1},E_{1},t}(\Theta)$ being equal to
zero.

The Master equation is coupled with the relative motion that the
excitation energy $E_{k}$ and the interaction time $\tau_{int}$
(The evolution time t is from $t=0$ to $t=\tau_{int}$) are
calculated by the parameterization method of the classical
deflection function \cite{Wol78,Li81} for each incident orbital
angular momentum.

\section{Results and Discussion}
\subsection{The driving potential of the DNS and the optimal excitation
energy to form SHN} From Eqs.(9-15), one finds that the nucleon
transition probability in Eq.(\ref{ma-quasi}) is related to the
size of the valance space $\Delta\varepsilon_{k}$ of
Eq.(\ref{11}), and so is related to the local excitation energy
$\varepsilon^{*}$, which is a function of the mass asymmetry of
the system via the driving potential of Eq.(\ref{17}) for a
certain angular momentum. Thus the driving potential is of vital
importance for the dynamical diffusion process. The calculated
driving potentials with and without considering the ground state
deformations of nuclei for the system $^{70}Zn +
^{208}Pb\rightarrow ^{278}112$ are shown in Fig.1 as a function of
the mass asymmetry variable $\eta$ in a bold solid line and thin
dashed line, respectively.  In the figure the ground state
deformation $\beta_{2}$ of the nuclei of the DNS is taken from
Ref. \cite{Moe95}. The bigger difference between the two lines
indicates the bigger deformation of nucleus. The orientation of
the deformed nuclei and the distance between the centers of the
two nuclei are taken in a way which gives the lowest
nucleus-nucleus interaction energy. Since the distribution
function $P(A_{1},E_{1},t)$ in Eq.(\ref{ma-quasi}) is considered
to cover the region from $A_{1}=0$ to $A_{1}=A_{P}+A_{T}=A$, the
driving potential has been calculated to cover $\eta=-1$ to 1. In
Fig.1 the arrow at $\eta_{i}$ points to the incident channel. One
nucleon transfer from $\eta_{i}$ to both sides, whether it is a
neutron or a proton, depends in which direction the potential
energy is lower. It turns out that the isotopic composition of the
nuclei forming the DNS determined in this way does not deviate
much from that following the condition of $N/Z$ equilibrium in the
system. Consequently, the driving potential of Eq.(17) is an
explicit function of neutron and proton numbers of fragments. In
order to form a compound nucleus, a barrier $B_{fus}^{*}$ shown in
the figure must be overcome. It is indicated that the deformation
of the nuclei decreases the potential energy and
 the inner fusion barrier a great deal. The
energy needed to pass over the barrier must be supplied by the
incident energy. The survival probability demands the lowest
excitation energy, so the optimal excitation energy of the
compound nucleus indicated in the figure is
$E_{CN}^{*}=U(\eta_{i})+B_{fus}^{*}$ where $U(\eta_{i})$ is the
potential energy of the initial DNS. For a set of cold fusion
reactions, the driving potentials are calculated and the obtained
optimal excitation energies of the compound nuclei from reactions
based on Pb target are shown in Fig.2 in open circles, and
compared with experimental data \cite{Hof88,Hof20} shown in solid
dots. Good agreement is found.

\subsection{The fusion probability $P_{CN}$}

Solving the Master equation Eq.(\ref{ma-quasi}) numerically, the
time evolution of $P(A_{k},E_{k},t)$ to find fragment k (mass
number $A_{k}$) with excitation energy $E_{k}$ at time t is
obtained. All the components on the left side of the fusion
barrier in Fig.1 contribute to the compound nuclear formation. The
fusion probability $P_{CN}$ is the summation  from $A_{1}=0$ to
$A_{BG}$:

  \begin{equation}\label{19}
  P_{CN}(J)=\int_{A_{1}=0}^{A_{BG}}P(A_{1},E_{k}(J),\tau_{int}(J))dA_{1}.
  \end{equation}
The intrinsic excitation energy $E_{k}$ is attributed to the
kinetic energy loss of the relative motion. The calculation of the
average energy loss, angular momentum loss, and interaction time
has been described in detail in Ref. \cite{Wol78}. Here the
relaxation times $\tau_{R}$ for radial kinetic energy,
$\tau_{\mathcal{J}}$ for angular momentum, and $\tau_{\epsilon}$
for spheroidal deformation have been determined as $\tau_{R}\simeq
0.3 \times 10^{-21}s$, $\tau_{\mathcal{J}}\simeq 1.5 \times
10^{-21}s$ and $\tau_{\epsilon}\simeq 4 \times 10^{-21}s$.  We
plot the dissipated kinetic energy and the mean interaction time
as a function of incident angular momentum J for $^{70}Zn+^{208}Pb
$ with an optimal kinetic energy $E_{cm}=E_{CN}^{*}(10.29
MeV)-Q=252.37MeV$ in Fig.3. The large energy damping below the
interaction barrier reflects the fragment deformation. The
reversible shape oscillations or other coherent modes of
excitation are not considered, only the effects of the
deformations which become irreversible due to the coupling with
the intrinsic degrees of freedom are taken into account
\cite{Wol79}. The radial kinetic energy, the relative and the
intrinsic angular momentum, the Coulomb interaction are all
affected by the nuclear deformations, and also the interaction
time $\tau_{int}(J)$.  For partial waves with small incident
angular momentum , the interaction time of the composite system is
very long, and during this time a large amount of kinetic energy
are dissipated and many nucleons are exchanged, some fraction of
the distribution probability contributes to compound nuclear
formation. The angular momentum dependence of $P_{CN}(J)$ for the
above mentioned case is shown in Fig.4(a), where the angular
momentum is cut off at about 20$\hbar$ because at larger angular
momentum the fission barrier for the compound nucleus becomes very
small. About $95\%$ fusion probability remain at $J=10\hbar$, and
about $68\%$ at $J=20\hbar$ with respect to $J=0$. The dissipated
energy, so as the excitation energy of nuclei are not influenced
much by the incident angular momentum up to $J=20\hbar$ as
indicated in Fig.3(a). But it may be found in Fig.3(b) that the
interaction time decreases rapidly with the increasing angular
momentum.
So the fusion probability $P_{CN}(J)$ decreases with the angular
momentum slowly.


 Fig.5(a) shows the calculated values of $P_{CN}$ for Pb-based
reactions at nearly central collisions $(J\sim 0)$ and with the
reaction energies according to those optimal excitation energies
indicated in Fig.2, respectively. Full dots are calculated results
by Eq.(\ref{ma-quasi}) without considering the quasi fission. Open
triangles are those including the quasi fission. One may find that
$P_{CN}$ with the consideration of the quasi fission decreases by
about four orders of magnitude with Z increasing from 106 to 118.
Because the inner fusion barrier $B^{*}_{fus}$ increases with
decreasing mass asymmetry of the initial DNS, i.e. with increasing
Z for the Pb-based reactions, the fusion probabilities decrease
rapidly with increasing Z. The straight line in the figure is used
to guide the eye. The consideration of the quasi fission process
in the master equation diminishes the fusion probability by one
order of magnitude for $Kr+Pb \rightarrow 118$. The decreasing
magnitude of the fusion probability  becomes less and less for
increasing asymmetry of the incident reaction system, since the
inner fusion barrier is getting decreasing, and the distribution
probability gets less chance to go to mass symmetrical direction,
to which the quasi fission barrier is getting smaller.

  \subsection{The survival probability of excited compound nucleus}
The super-heavy compound nuclei are formed in excited states, and
will lose excitation energy mainly by  emission of particles and
$\gamma$ quanta, and by fission. The surviving probability in cold
fusion reactions estimates the competition between fission and
neutron evaporation in the excited compound nucleus by a
statistical model.  In this case the width for the emission of a
charged particle is much less than that for the emission of a
neutron, and the $\gamma$ ray emission is important only when the
excitation energy is smaller than the one-neutron separation
energy.

In cold fusion reactions, the survival probability under
one-neutron emission can be written as:

\begin{equation}\label{20}
  W_{sur}(E^{*}_{CN},J)=P_{1}(E^{*}_{CN},J)\frac{\Gamma_{n}(E^{*}_{CN},J)}
  {\Gamma_{n}(E^{*}_{CN},J)+\Gamma_{f}(E^{*}_{CN},J)},
  \end{equation}
where $E^{*}_{CN}$, and $J$ are the excitation energy and the
angular momentum of the compound nucleus, respectively.
$P_{1}(E^{*}_{CN},J)$ is the realization probability of the 1n
channel at given $E^{*}_{CN}$ and $J$, which is calculated with
the expression of Eq.(7) from Ref. \cite{Adac00}. $\Gamma_{n}$ and
$\Gamma_{f}$ are the widths of neutron emission and fission,
respectively.  In calculating $W_{sur}$ the following formulae are
used:

\begin{equation}\label{21}
  \Gamma_{n}(E^{*})=\frac{1}{2\pi\rho(E^{*})}\cdot\frac{2M_{n}R^{2}}{\hbar^{2}}g
  \int_{0}^{E^{*}-B_{n}-1/a}\varepsilon
  \rho(E^{*}-B_{n}-\varepsilon)d\varepsilon,
  \end{equation}
and

  \begin{equation}\label{22}
  \Gamma_{f}(E^{*})=\frac{1}{2\pi\rho(E^{*})}\int_{0}^{E^{*}-B_{f}-1/a}
  \rho(E^{*}-B_{f}-\varepsilon)d\varepsilon,
  \end{equation}
where $\rho(E^{*})=\frac{1}{\sqrt{48}E^{*}}exp[2\sqrt{aE^{*}}]$ is
the level density, R ,$B_{f}$, $B_{n}$ are the radius, the fission
barrier and the neutron separation energy of the compound nucleus,
respectively. $M_{n}$ is the mass of the neutron, $g$ the spin
factor of neutron and  $a$  the  level density parameter which is
taken to be $a=A/12$. $E^*$ is the effective excitation energy of
the compound nucleus.  In Ref. \cite{Adac00} the fission barrier
for SHE is divided into the macroscopic part $B^{LD}_{f}$,
determined by liquid-drop model, and into the microscopic part
$B^{Mic}_{f}$, determined by shell correction. The microscopic
energy will be damped due to the dependence of the shell effects
on the nuclear excitation. Thus, the fission barrier can be
written as :

\begin{equation}\label{23}
  B_{f}=B^{LD}_{f}+B^{Mic}_{f}(E^{*}=0)exp[-\frac{E^{*}}{E_{D}}]
  -(\frac{\hbar^2}{2J_{g.s.}}-\frac{\hbar^2}{2J_{s.d.}})J(J+1),
  \end{equation}
where $E_{D}$ is a damping factor describing the decrease of the
influence of the shell effects on the level density with the increasing
excitation energy of the nucleus, which is taken as
\begin{equation}\label{24}
  E_{D}=0.4A^{4/3}/a,
  \end{equation}
where $A$ is the mass number of the nucleus.
$J_{g.s.;s.d.}=k\frac{2}{5}MR^2(1+\beta_2^{g.s.;s.d.}/3)$ are the
moment of inertia of the fissioning nucleus at its ground state
and the saddle state, respectively. Where $k \approx 0.4$
\cite{Zagc01}.  Since there are no data available, the quadrupole
deformation parameters $\beta_2$ at the saddle point are taken
from the microscopic calculation of the relativistic mean field
(RMF) theory \cite{Zha03}, which has been proven to be quite
successful for the description for exotic nuclei and superheavy
nuclei (SHN) \cite{Men96,Men98,Menl98}. The angular momentum
dependence of $W_{sur}$ for mentioned reaction
$^{70}Zn+^{208}Pb\rightarrow ^{278}112$ is indicated in the
Fig.4(b). One may find that at $J\sim 15\hbar$ the survival
probability decreases about one order of magnitude.

The macroscopic fission barrier can be evaluated by liquid drop
model \cite{Van73}.
Taking the neutron separation energy and $B^{Mic}_{f}$ from
Ref. \cite{Smo99}, the calculated survival probabilities for
one-neutron emission Pb-based reactions at nearly central
collisions with the excitation energies from Fig.2 are shown in
Fig.5(b). The tendency of the results is basically consistent with
that shown in Fig.4 of Ref. \cite{Adac00}.

\subsection{The evaporation residue cross section}
Applying Eqs.(1) and (2) we calculated the evaporation residue
cross sections. For cold fusion reactions with the optimal
excitation energies as indicated in Fig.2, a set of evaporation
residue cross sections for Pb-based reactions are shown in Fig.6
with solid stars, the solid dots are experimental data quoted in
Ref. \cite{Adan00}, and some estimated data for element 114, 116,
and 118 by different groups are indicated in the figure
\cite{Den00,Smo01,Ada01}. The upper limit for element 118 was
estimated by LBNL recently \cite{Gre03} and also shown in this
figure. Our results are in principle in agreement with the data
within one order of magnitude.

\section{Summary}

The fusion probability is calculated in very strongly damped
reaction processes, where large amounts of the relative kinetic
energy are changed into intrinsic excitation energy and nucleons
transferred from the lighter fragment to the heavier one to
produce super-heavy nuclei in the tail of the heavy-fragment mass
distribution. Within the DNS conception, instead of solving FPE
analytically, the Master equation is solved numerically in order
to calculate the fusion probability, so that the harmonic
oscillator approximation to the potential energy of the DNS, which
is the very entrance of the nuclear structure of the
model \cite{Ada97,Ada98}, is avoided. In our calculations the
relative motion including the relaxations of the energy, angular
momentum, and fragment deformation is explicitly treated, so that
the nucleon transition probabilities, which are derived
microscopically, are coupled with the relative motion and thus are
time-dependent. Comparing with the analytical (or the
logistic-type) solution of FPE, our results preserve more
dynamical effects. The fusion process is calculated for each
partial wave, and an about $32\%$ fusion probability reduction is
found for the $J=20$ partial wave compared with that for the
central collision to form compound nucleus 112. And  the survival
probability at $J\sim 15\hbar$ decreases about one order of
magnitude.  Our calculated evaporation residue cross sections for
one-neutron emission channel of Pb-based reactions are basically
in agreement with the experimental data within one order of
magnitude. However, although the driving potential has been
calculated in $\eta$ and R two dimensions, the diffusion process
to the two dimensions are not treated simultaneously. The quasi
fission is treated by a decay rate of Kramers' type.  The Master
equation should be extended into a two dimensional case by taking
the distance between the centers of nuclei into account in
addition, so that the qusi fission could be described in the
process to fully understand the reaction dynamics.  Presently, the
nucleus-nucleus interaction with deformations is only simulated by
shifting the distance between surface of spherical nuclei to a
smaller relative distance determined by the same distance as those
which the deformed nuclei have, and in this way it has been a
little overestimated, especially in the region where nucleus has
bigger deformation. Therefore, the nucleus-nucleus interaction
including the consideration of nuclear deformation is being
investigated by using the Skyrme-type force, which we will include
in the calculation later, and would like to publish
elsewhere \cite{Jia05}. In future a time-dependent multidimensional
potential energy surface has to be built up, and to treat the
time-dependent dynamics in order to get a complete quantitative
understanding of the fusion reaction mechanism of heavy nuclei.

{\section{ACKNOWLEDGMENTS}}

The work is supported by the National Natural Science Foundation
of China under Grant No.10175082, 10235020, 10375001, 10311130175;
the Major Basic Research Development Program under Grant No.
G2000-0774-07;  the Knowledge Innovation Project of the Chinese
Academy of Sciences under Grant No. KJCX2-SW-N02; One Hundred
Person Project of CSA;  the CASK.C. Wong Post-doctors Research
Award Fund; the Alexander von Humboldt-Stiftung of Germany; the National key program for Basic Research of the
Ministry of Science and Technology (2001CCB01200, 2002CCB00200);
Natural Science Foundation of Guangdong province 04300874, and the
financial support from DFG of Germany.

\newpage
{\large{{\bf Figure Captions:}}}

 Fig.1: The driving potential of the DNS for the system $^{70}Zn
+ ^{208}Pb\rightarrow ^{278}112$  as a function of the mass
asymmetry variable $\eta$. $BG$ marks the top point of the
potential energy.

Fig2: The optimal excitation energies of the compound nuclei from
reactions  based on Pb target as a function of the charge number
of compound nuclei. The calculated results are shown with open
circles, and the experimental data with solid dots.

Fig3: The mean dissipated kinetic energy and the mean interaction
time of the relative motion are shown as a function of the
incident angular momentum J in (a) and (b), respectively, for
$^{70}Zn+^{208}Pb $ reaction with the corresponding optimal
excitation energy $E_{CN}^{*}=10.29 MeV$ of the compound nucleus
$^{278}112$.

Fig4: (a): The angular momentum dependence of $P_{CN}(J)$ for the
same case as in Fig.3. (b): The corresponding angular momentum
dependence of $W_{sur}$.

Fig5: (a): the calculated values of the fusion probability
$P_{CN}$ for one-neutron emission Pb-based reactions at nearly
central collisions  and with the reaction energies according to
those indicated in Fig.2 as a function of the charge number of the
compound nuclei. The open triangles and solid dots stand for the
fusion probability $P_{CN}$ with and without considering the
effect of the quasi fission, respectively.  The corresponding mass
number are listed on the second row. (b): The corresponding
calculated survival probability.

Fig6: The evaporation residue cross sections for one-neutron
emission Pb-based reactions with the excitation energies from
Fig.2  as a function of the charge number of compound nuclei.  Our
calculated results are indicated by solid stars, the experimental
data by solid dots. And some estimated data for element 114, 116,
and 118 by different groups are indicated with different symbols.


\bigskip

\begin{thebibliography}{99}

\bibitem{Nil68}S. G. Nilsson, J. R. Nix, A. Sobiczewski, Z. Szymanski, S. Wycech, C. Gustafson,
                and P. Moeller, Nucl. Phys. {\bf A115},
                545(1968).
\bibitem{Mel67}H. Meldner,1967, in Proceedings of the Int. Symp. on Nuclides far off the Stability Line,
W. Forsling, C. J. Herrlander, and H. Ryde, Eds., Lysekil, Sweden,
August 21-27, 1966, Ark. Fys. {\bf 36}, 593.
\bibitem{Mos69}U.Mosel and W. Greiner, Z. Phys. {\bf A222},
61(1969).
\bibitem{Fis72}E. O. Fiset and J. R. Nix, Nucl. Phys. {\bf A193},
647(1972).
\bibitem{Ran76}J. Randrup, S. E. Larsson, P. Moeller, S. G. Nilsson, K. Pomorski, and A. Sobiczewski,
               1976 Phys. Rev.  {\bf C13}, 229(1976).
\bibitem{Hof88}S. Hofmann, Rep. Prog. Phys. {\bf 61},
636-689(1998).
\bibitem{Oga99}Yu. Ts. Oganessian et al., Nature {\bf 400},
242-245(1999).
\bibitem{Oga001}Yu. Ts. Oganessian et al., Phys.Rev.C{\bf 62}, 041604(R)(2000),
   Phys.Rev.C{\bf 613},
011301(R)(2000).
\bibitem{Hof20}S.Hofmann and G.Muenzenberg, Rev. Mod. Phys. 72,
733(2000).
\bibitem{Smo99}Robert Smolanczuk  Phys.Rev.C{\bf59},
2634-2639(1999).
\bibitem{She02}Caiwan Shen, Grigori Kosenko, and Yasuhisa Abe, Phys.Rev.C{\bf 59},
061602(R)(2002).
\bibitem{Fro84}P. Froebrich, Phys. Rep. {\bf 116},
337-400(1984).
\bibitem{Zag01}V.L.Zagrebaev, Phys.Rev.C{\bf64}, 034606(2001).
\bibitem{Che96}E.A. Cherepanov, V.V. Volkov, N.V.Antonenko, V.B.Permjakov and
A.V. Nasirov,  Nucl. Phys. A{\bf 459} (1996)145.
\bibitem{Ada97}G.G.Adamian, N.V.Antonenko, W.Scheid, Nucl.Phys. A{\bf 618},
176-198(1997).
\bibitem{Ada98}G.G.Adamian, N.V.Antonenko, W.Scheid, V.V. Volkov, Nucl.Phys. A{\bf 633},
409-420(1998).
\bibitem{Ant93}N.V.Antonenko, E.A. Cherepanov, A.V. Nasirov, V.B.Permjakov
  and V.V.Volkov,
 Phys.Lett.B{\bf 319}, 425(1993); Phys.Rev.C{\bf 51}, 2634(1995).
\bibitem{Ada972}G.G.Adamian, N.V.Antonenko, W.Scheid, V.V. Volkov,
   Nucl.Phys. A{\bf 627}, 361(1997).
\bibitem{Adac03}G. G. Adamian, N. V. Antonenko, and W. Scheid,
 Phys. Rev. C {\bf 68}, 034601(2003).
\bibitem{Dia01}A. Diaz-Torres, G. G. Adamian, N. V. Antonenko, and W. Scheid,
 Phys. Rev. C{\bf 64}, 024604(2001).
 \bibitem{Wol78}G. Wolschin and W.N\"{o}renberg, Z. Phys. A{\bf284},
209-216(1978).
\bibitem{LWF03}Li W F, Wang N, Li J F et al.,  Europhys. Lett. {\bf 64}, 750(2003)
\bibitem{Nor74}W. N\"{o}renberg, Phys.Lett.B{\bf 53},
289(1974).
\bibitem{Mor75}L.G.Moretto and J.S. Swentek, Phys.Lett.B{\bf58},
26(1975).
\bibitem{Hof76}H.Hofmann and P.J.Siemens, Nucl.Phys.A{\bf 257}, 165(1976); A{\bf 275},
464(1977).
\bibitem{Ran78}J.Randrup, Nucl.Phys.A{\bf 307}, 319(1978); A{\bf327},
490(1979).
\bibitem{Wei80}H.A.Weidenm\"{u}ller, Prog.Part.Nucl.Phys.{\bf 3},
49(1980).
\bibitem{Fro80}P.Froebrich, B. Strack and M. Durand, Nucl.Phys.A{\bf406}, 49(1980);
P.Froebrich and S.Y.Xu, Nucl.Phys.A{\bf 477}, 143(1988).
\bibitem{Fel87}H.Feldmeier, Rep.Prog.Phys. {\bf 50}, 1(1987).
\bibitem{Li88}Li Junqing, Liu Jianye, Zhu Yongtai, Zhu Jieding, High
Energy Phys. and Nucl. Phys. 12(1988)189.
\bibitem{Ant92}N.V.Antonenko and R.V.Jolos, Z.Phys.A{\bf341},
459(1992).
\bibitem{Ada94}G.G.Adamian, N.V.Antonenko, R.V.Jolos and A.K.Nasirov, Phys. Part. Nucl.{\bf
25},583(1994); Nucl. Phys. A{\bf551}, 321(1993).
\bibitem{Ant94}N.V.Antonenko, S.P.Ivanova, R.V.Jolos and W.Scheid, Phys.Rev.C{\bf50},
2063(1994).
\bibitem{Ayi76}S.Ayik, B.Schuermann and W.N\"{o}renberg, Z.Phys.A{\bf277}, 299(1976); Z.Phys.A{\bf279}, 145(1976);
B.Schuermann,W.N\"{o}renberg and M.Simbel, Z.Phys.A{\bf286},
263(1978).
\bibitem{Fre71}J.B.French, K.F.Ratcliff, Phys.Rev.C3, 94(1971).
\bibitem{Cha71}F.S.Chang, J.B.French and T.H.Thio, Ann.Physics(N.Y.)66,
137(1971).
\bibitem{Ayi74}S.Ayik, J.N.Ginocchio, Nucl.Phys.A221, 285(1974); Nucl.Phys.A234,
13(1974).
\bibitem{Li81}J.Q.Li, X.T.Tang and G.Wolschin, Phys.Lett.B{\bf 105}, 107(1981);
J.Q.Li and G.Wolschin, Phys.Rev.C{\bf 27}, 590(1983).
\bibitem{Moe95}P. M\"oller, J.R. Nix, W.D. Myers, and W.J.
Swiatecki, At. Data Nucl. Data Tables, {\bf 59}, 185(1995).
\bibitem{Wol79}G.Wolschin , Phys. Lett.{\bf B88}, 35(1979).
\bibitem{Adac00}G.G. Adamian, N.V.Antonenko, S.P.Ivanova, W.Scheid,
            Phys.Rev.C62, 064303 (2000).
\bibitem{Zagc01}V. I. Zagrebaev, Y. Aritomo, M. G. Itkis, and Yu. Ts. Oganessian,
      Phys. Rev. {\bf C65},014607(2001).
\bibitem{Zha03}Zhang W, Zhang S S, Zhang S Q, and Meng J, Chin. Phys. Lett {\bf 20},
           1694(2003).
\bibitem{Men96}Meng J, and Ring P, Phys. Rev. Lett. {\bf
77},3963(1996).
\bibitem{Men98}Meng J, Nucl. Phys. {\bf A635},3(1998).
\bibitem{Menl98}Meng J, and Ring P, Phys. Rev. Lett. {\bf 80},
460(1998).
\bibitem{Van73}R.Vandenbosch, J.Huizenga, Nuclear Fission,
Academic Press, New York, London (1973).
\bibitem{Adan00}G.G. Adamian, N.V.Antonenko, W.Scheid, Nucl.Phys.A678,
24-38(2000).
\bibitem{Den00}V.Yu. Denisov, S. Hofmann, Phys. Rev. C {\bf 61},
034606(2000).
\bibitem{Smo01}R. Smola$\acute{n}$czuk, Phys. Rev. C {\bf 63},
044607 (2001).
\bibitem{Ada01}G.G. Adamian, N.V. Antonenko, A. Diaz-Torres, W.
Scheid, Yu.M. Tchuvil'sky, AIP Conf. Proc. {\bf 561}, 421(2001).
\bibitem{Gre03}K.E. Gregorich, {\it et al.}, Eur. Phys. J. A {\bf
18}, 633(2003).
\bibitem{Jia05}Jia Fei, Xu Hushan, Huang Tianheng {\it et al.}, Chin. Phys. Lett. {\bf
22}, 1374(2005).
\end{thebibliography}
\end{document}